\documentclass{IEEEcsmagArxiv}

\usepackage[colorlinks,urlcolor=blue,linkcolor=blue,citecolor=blue]{hyperref}

\usepackage{upmath}

\usepackage{subfig}
\usepackage{multirow}
\usepackage{comment}
\usepackage[11pt]{moresize}
\usepackage{xcolor}

\definecolor{color2017Survey}{HTML}{b3cde3}
\definecolor{color2018Survey}{HTML}{ccebc5}
\definecolor{color2019Survey}{HTML}{FFB1B1}

\definecolor{colorFamil}{HTML}{548235}
\definecolor{colorFound}{HTML}{7030A0}
\definecolor{colorSkill}{HTML}{C55A11}
\definecolor{colorSyste}{HTML}{FFC000}
\definecolor{colorPeerR}{HTML}{7F7F7F}
\definecolor{colorProje}{HTML}{F2F2F2}

\newcommand{\ColorCode}{\setlength{\fboxrule}{0.5pt}\setlength\fboxsep{3pt}
Color-coding: 
    \fcolorbox{black}{color2017Survey}{\color{color2017Survey}\tiny a}~2017~/ 
    \fcolorbox{black}{color2018Survey}{\color{color2018Survey}\tiny a}~2018~/ 
    \fcolorbox{black}{color2019Survey}{\color{color2019Survey}\tiny a}~2019}

\newcommand{\fromReviews}[1]{\textcolor{blue}{\textbf{\textit{#1}}}}

\begin{document}

\title{Through the Looking Glass: Insights into Visualization Pedagogy through Sentiment Analysis of Peer Review Text}

\author{Zachariah Beasley}
\affil{University of South Florida}

\author{Alon Friedman}
\affil{University of South Florida}

\author{Paul Rosen}
\affil{University of South Florida}


\begin{abstract}
Peer review is a widely utilized feedback mechanism for engaging students. As a pedagogical method, it has been shown to improve educational outcomes, but we have found limited empirical measurement of peer review in visualization courses. In addition to increasing engagement, peer review provides diverse feedback and reinforces recently-learned course concepts through critical evaluation of others' work. We discuss the construction and application of peer review in two visualization courses from different colleges at the University of South Florida. We then analyze student projects and peer review text via sentiment analysis to infer insights for visualization educators, including the focus of course content, engagement across student groups, student mastery of concepts, course trends over time, and expert intervention effectiveness. Finally, we provide suggestions for adapting peer review to other visualization courses to engage students and increase instructor understanding of the peer review process.
\end{abstract}




\maketitle

\chapterinitial{In recent years}, visual communication has shown growth in many professional fields and media, from scientific discovery to humanities communications. It is thus no surprise that we are experiencing dramatic growth in the number of students enrolling in visual communication courses---a growth that makes it difficult to provide the detailed subjective feedback students need to improve the quality of their work~\cite{friedman2018developing}. The recent proliferation of online education has further increased this pressure. 

While visualization education as a whole remains a work in progress~\cite{domik2000we}, \cite{rushmeier2007revisiting},  we have identified two broad visualization education themes. The first is the proper construction of visualizations---teaching students the algorithms and visualization principles to use in creating visualizations. The second is the subjective evaluation of the quality and accuracy of visualizations. In other words, training students to \textit{critically evaluate others' visualizations}, which is the focus of our work.  These necessary skills are commonly taught through informal methods, such as group or whole-class discussions. However, informal methods which lack active participation leave students' skills underdeveloped~\cite{santos2018heuristic}.   Isenberg et al.\ expanded upon this understanding by emphasizing ``collaborative visualization''~\cite{isenberg2011collaborative}, further defined as the contributions of different experts toward a shared goal of understanding the visual object, phenomenon, or data under investigation~\cite{raje1998cev}.


Peer review is a highly-engaging feedback mechanism often associated with liberal arts courses but also found in professional code review and scholarly publication. It is particularly suited to visualization education~\cite{beasley2020leveraging, Friedman2017}. Rather than rely solely on instructors for visualization feedback, peers collaborate to provide diverse, multi-sourced feedback to one another. Most importantly, the \textit{evaluation process itself gives students an opportunity to reinforce recently-learned course concepts through formal critical evaluation}.

In this article, we take a deeper look at four years of insights gained from peer review in visualization education. We begin by describing the two peer review-oriented visualization courses we use in our analysis---one from the field of Computer Science and the other from the field of Mass Communications. This includes a discussion on our generic peer review rubric, which was customized to reinforce key concepts in each course. Next, we describe the collection of peer review text from multiple semesters of each course and the evaluation of the data using sentiment analysis and aspect extraction. Rather than directly compare the distinct courses, our primary goal is to demonstrate the type of information that can be obtained from peer review text, which is highly dependent on the visualization course context and data collected. This methodology allows an instructor to more thoroughly analyze the peer review process to 1) determine its effectiveness at engaging students, 2) understand whether students have mastered course concepts, 3) note course trends over time, and 4) identify the effects of expert intervention. Finally, we share insights that others may draw from to improve their own visualization courses through the use of peer review.

\section{PEER REVIEW IN VISUALIZATION}



In a review of several major visualization venues, InfoVis, SciVis, VAST, EuroVis, and Pacific Vis, we found no empirical evaluation of students' work or engagement in the classroom, aside from our own~\cite{beasley2020leveraging, Friedman2017}. The use of peer review in the visualization classroom thus remains an open area of research, despite the fact that critiquing has long been acknowledged as a critical part of the visualization design process~\cite{kosara2007visualization}. 

Fortunately, other disciplines have reported extensively on peer review, and many of their insights transfer to the visualization education domain. In the liberal arts, e.g., numerous researchers have investigated whether the ability to write can be mastered in one context and easily transferred to another, a claim that remains disputed~\cite{wardle2012addressing}.
However, this concern primarily impacts the use of \textit{summative} reviews for assessment purposes (i.e., providing a grade). It does not diminish the value of using \textit{formative} peer review to stimulate learning. In fact, peer review has been shown to be a highly engaging, active-learning mechanism~\cite{weaver2012peer}. Active-learning, in general contexts, has been shown to improve comprehension, retention, and overall educational outcomes over non-active-learning approaches~\cite{doi:10.1152/advan.00053.2006}.

\section{A THEORETICAL FRAMEWORK FOR PEER REVIEW IN VISUALIZATION}

Education researchers constantly rely on theoretical frameworks to explain student behavior, experience, and means of success. Many of those educational theories are designed to enhance students' learning experience when teaching takes place. In the field of visualization education, Oliver \& Aczel reviewed four major education theories, including cognitive theory, cognitive dimension, Popper-Campbell model, and Vygotsky's theory~\cite{oliver2002theoretical}. They evaluated their translation to visualization education under logic learning processes, where students acquire logic steps to review a visualization.

In the field of writing analytics, researchers often cite Vygotsky's model for its influence on student peer review engagement success. Moxley notes that Vygotsky's theory helps instructors to improve student reading, writing, and collaboration skills using an online dashboard dedicated to peer review~\cite{moxley2010afterword}. Using Vygotsky's theory, Moxley built a rubric to accommodate student engagement.  Our previously reported peer review model for visualization was influenced by Moxley's rubric and Vygotsky's theory~\cite{beasley2020leveraging,Friedman2017}.

In the field of visualization education, Oliver \& Aczel examined Vygotsky's model of the zone of proximal development (ZPD)~\cite{oliver2002theoretical}. The ZPD model focuses on three important components which aid the learning process: 1)~the presence of an individual with knowledge and skills beyond that of the learner, 2)~social interactions with a tutor that allows the learner to practice skills, and 3)~supportive activities provided by the educator or more competent peers. Oliver \& Aczel reported that the ZPD model could provide active and tailored feedback to the learner, supporting the student's learning experience.
Using peer review in visualization courses supports the social and peer components of the ZPD model.

\section{A PEER REVIEW RUBRIC FOR VISUALIZATION}
\label{sec.rubric}

Rubrics are used by instructors in a variety of disciplines to provide feedback or to grade student products, e.g., writings, presentations, and portfolios~\cite{beasley2019designing}. Generally speaking, a rubric is a pedagogical tool that articulates the expectations for an assignment. It lists the most important criteria and describes levels of competency from poor to excellent. 

However, effective rubrics are difficult to synthesize from scratch---they must consider core concepts of the course, project requirements, and overall educational goals. 
In our Pedagogy of Visualization 2017 workshop paper, we addressed the limited availability of rubrics for peer review in the visualization classroom by providing a generic rubric that can be customized to the needs of the individual instructor, course, and project~\cite{Friedman2017}. We later evaluated the efficacy of the rubric in a computer science visualization course~\cite{beasley2020leveraging}. A \LaTeX~version of the rubric can be retrieved at \url{https://github.com/USFDataVisualization/VisPeerReview}.

The rubric was built by reviewing the content of multiple visualization courses and extracting key concepts necessary for demonstrating proficiency in learning their objectives. The structure divides the rubric into five major assessment categories: algorithm design, interaction design, visual design, design consideration, and visualization narrative. The algorithm design category focuses on algorithm selection and implementation. Interaction design concentrates on how the user interacts with the visualization. Visual design relates to the technical aspects of how data are placed in the visualization (e.g., the expressiveness and effectiveness of visual encoding channels). Design consideration corresponds to the composition and aesthetic aspects of the visualization, e.g., embellishments. Visualization narrative is the final category, which is used in situations where the story is as important as the visualization itself.

Each of the five major categories has a variety of assessments associated with it. For scoring purposes, each assessment is affixed to a 5-point scale. Furthermore, the intent goes beyond numeric scores---providing a text-based comment is encouraged for every assessment category. One final key element of our original rubric design was the intention for a high-level of customization based upon the content of a course or project. As such, the rubric is versatile enough to work with courses that have very different focuses, which we demonstrate in this article.

\section{COURSES OVERVIEW}
\label{sec.course}

\subsection{Data Visualization Course}

Our first course entitled ``Data Visualization'' was taught in the Computer Science department in the College of Engineering at the University of South Florida. The course was offered as a mixed undergraduate/graduate elective.

\textbf{Course Content.}
The course was taught using Munzner's \textit{Visualization Analysis \& Design}~\cite{munzner2014visualization} with additional research papers and outside visualization content (e.g., Vis Lies, New York Times Graphics, etc.) to generate discussions. 
The course materials were primarily presented  using lecture, research paper presentations by students, and discussions (e.g., small group and whole class critiques).


The main learning objectives for the course were that students would demonstrate the ability to: evaluate data and user requirements and program an effective visualization to match those requirements; associate elements of visualizations with the basic components, e.g., data abstractions and visual encodings, that are used in their construction; and critique interactive visualizations with respect to their effectiveness for selected tasks, visual encodings, and interaction design and implementation.

\textbf{Project Design \& Peer Review.}
We pursued structured assessment (i.e., well-defined projects), 
consisting of eight projects, one in Tableau and seven in Processing, totaling $50\%$ of the students' final grade. Projects deadlines were every 10-14 days, except for Project~6, which was assigned over Spring Break and allowed approximately 30 days. 
%
When designing projects, we desired that students gain visualization skills by demonstrating proficiency in using software engineering problem-solving techniques~\cite{cleveland1993visualizing}. 
Many projects included direct reuse of components and associated feedback from previous projects.

The project design included a dual goal of teaching visualization practice and software design. Therefore, the projects were divided into four categories:
\textit{Familiarization}: One project emphasized familiarizing students with using standard visualization types. 
\textit{Foundations}: These three projects emphasized the implementation of the basic mechanisms of visualization, e.g., data abstraction, visual encoding, and interaction.
\textit{Transferability}: In two projects, students applied their foundational skills to develop more complex visualization types.
\textit{Software Engineering}: These two projects use software engineering skills to build and enhance a visualization dashboard.


Projects were set up to maximally build upon and reuse components from previous projects while still challenging students with new project requirements. Therefore, the peer reviews serve as an integral feedback mechanism by allowing students to use feedback to refine the work they submit.
To build the peer review rubric, we customized the rubric per project. To do this, we simply extracted the relevant components from the full rubric template described earlier. For example, Project~1 was the only project that included a narrative component. Projects 4-8 required interaction, while Projects 1-3 did not. Finally, the sub-assessments included for early projects reflected only topics that had been covered in class to that point.

After each project, students provided reviews to three randomly selected peers' work using the provided rubric within 5-7 days. The peer review form consisted of a series of assessments, each scored on a 5-point Likert scale. Each assessment also had an optional input box for text-based feedback. Peer feedback was the primary form of qualitative feedback given to students. In addition, students could request additional feedback from the instructor. Students received small amounts of (spot checked) \textit{completion-based} credit, which was approximately $10\%$ of the final course grade. Projects were graded by the instructor and teaching assistants primarily using objective requirements, as well as some subjective judgment. Peer review scores did not influence the project grade.

%

\begin{figure*}[!t]
    \centering
    {\includegraphics[width=1\linewidth]{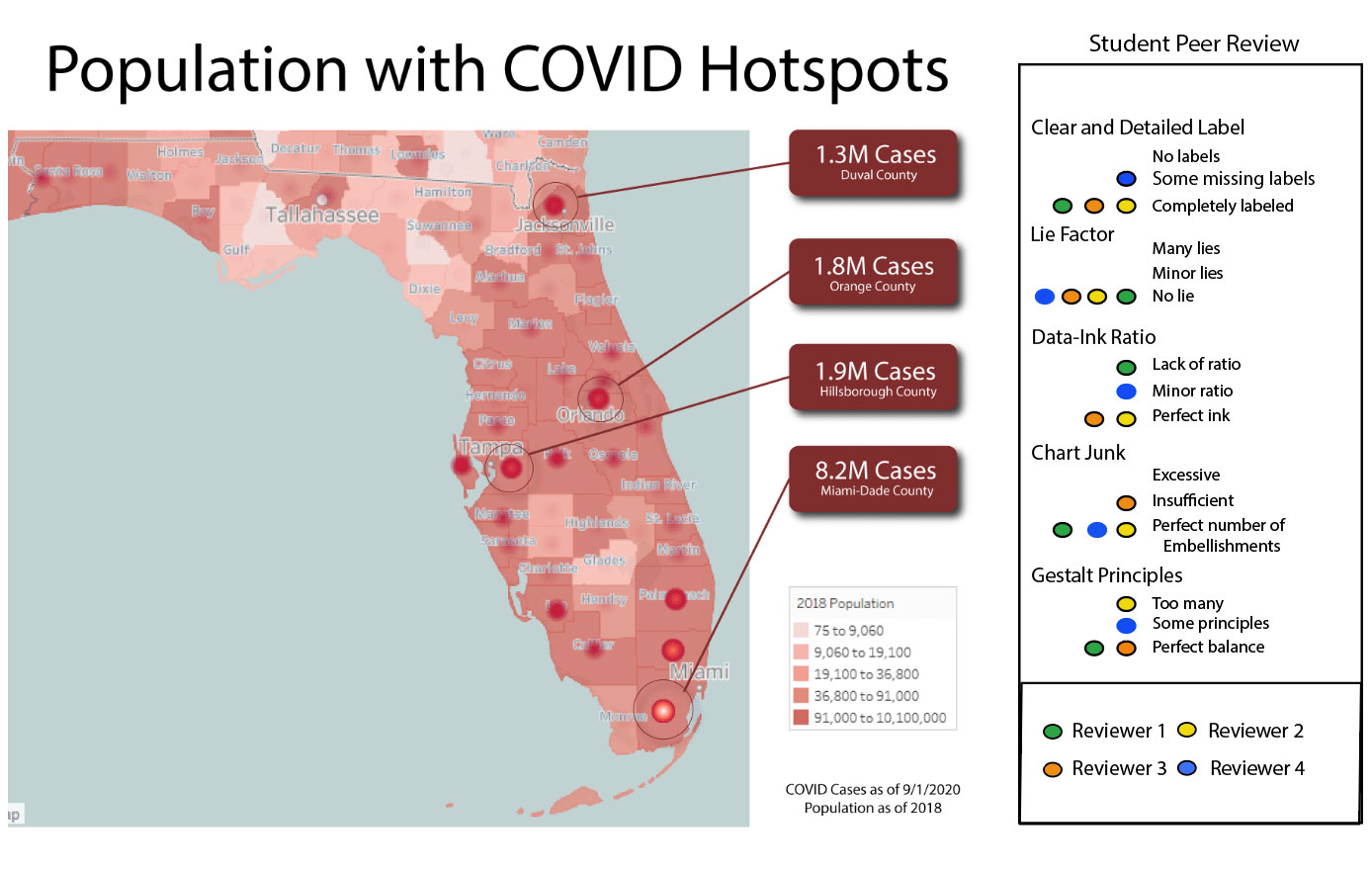}}
    \caption{Left: Submission for Visual Literacy visual map narrative. Right: Visualization of the qualitative feedback received from four peers.}
    \label{fig:visMapNarrative}
\end{figure*}

\subsection{Visual Literacy Course}
The second course, ``Introduction to Visual Communication," later renamed ``Visual Literacy," 
was taught in the Mass Communications department in the College of Arts and Sciences at the University of South Florida. The course was originally designed for mass communication and journalism undergraduate students. With a growing number of students enrolled in the course, it eventually became a required course for all undergraduate students enrolled in the College of Arts and Sciences, and the course content was expanded to provide freshmen with a stronger foundation for visual communication.

\textbf{Course Content.} Visual Literacy was a survey course designed to give students a basic knowledge of the many forms visual communication can take and provide practical experience in developing projects. It spanned a wide range of areas, including visual persuasion, data visualization, and visual storytelling. 
Under visual literacy course learning objectives, 
students would not only demonstrate the ability to use and generate graphic designs with design principles covered in the class but would also review their peers' work based on the designated rubric. 
Furthermore, continued moves toward social media analytics pushed our curriculum toward exploring various forms of visual content found on social and digital media.

\textbf{Project Design \& Peer Review.}
To gain visual communication skills, the course was based on applying the aesthetics associated with media productions and data findings. Assignments were divided into fourteen different modules, where each module addressed a different visual tool or software from using mobile phones as cameras to generating a logo with Adobe Creative Cloud to finding data and converting it to a visual map narrative. In the data visualization module, the students learned about data collection and utilized a combination of Plot.ly and Adobe Creative Cloud to generate infographics (see \autoref{fig:visMapNarrative}).


Students' performance was assessed via weekly assignments with peer reviews. In the data visualization module, the rubric that was utilized highlighted five major topics that matched the curriculum: clear detailed labeling, the lie factor, the data-ink ratio, chart junk data density, and Gestalt principles. The first category asked the student to search his/her peer's visualization for clear and detailed labeling of every aspect of the data represented in the graph or chart. The next three categories were based on Tufte's principles of design~\cite{tufte1990envisioning}. The lie factor directs the relationship between the size of the effect shown in the graphic and the size of the effect shown in the data. The data-ink ratio, the proportion of ink (or pixels) used to present data compared to the total ink used in the display, refers to the non-erasable ink used for the presentation of data. If data-ink were removed from the image, the graphic would lose its content. Accordingly, non-data-ink is used for scales, labels, and edges found in the visual work. The next category, known as chart junk, refers to all of the visual elements in charts that are not needed for comprehending the information on the chart or that distract from this information. Finally, Gestalt principles describe how the human eye perceives visual elements---in particular, they tell us that complex images reduce to simpler shapes.

In the data visualization assignment, each student was asked to create a map based on data found using leading open-source data repositories and visual dashboards. In the second stage of this assignment, each student was asked to evaluate each other's work using a peer review rubric. 
The peer review form consisted of a series of assessments, each scored on a 4-point Likert scale. In addition, a single input box was provided for text-based feedback. 
Each week students submitted their assignments and conducted peer reviews on five peers' assignments. At the same time, highly engaged students could optionally request detailed instructor feedback beyond peer reviews.
The overall grades in the class were set by the instructor, with the student's peer review grades counting for 10\% of the overall grade assigned.

\section{DATA COLLECTION}

To analyze both visualization courses, we collected significant quantities of data. We utilized quantitative analysis via natural language processing (specifically, sentiment analysis), coupled with qualitative analysis from representative student works. 

\subsection{Data Description}
\label{sec.course.data}

In the Data Visualization course, the open-ended review comments were gathered from eight projects in 2017 and seven projects in 2018 and 2019. The rubric contained multiple comment sections, each of which was concatenated into a single string to produce 3,116 reviews. Approximately 27\% of the comments (or 846 out of 3,116) had no sentiment keywords and could not be scored but still contributed to the part-of-speech analysis. Students were incentivized with a completion point (spot-checked by the teaching assistant) to submit a high-quality review. The numeric score from the rubric was not included in the dataset for comparison with the text. Finally, we manually reviewed student projects to find those which, in combination with peer comments, exemplified the value of peer review.


In the Visual Literacy course, we collected data from Spring 2017 through Fall 2020 (eight semesters). Open-ended peer review comments, with an associated numeric score, were collected from a single project. Students received a completion score for filling out a quality peer review. There were 1,571 reviews after those with no comment, no score, or an out-of-bounds score were excluded. Of the remaining reviews, 29\% (or 455 out of 1,571) had no sentiment keywords and could not be scored but still contributed to the part-of-speech analysis. 


\subsection{Analysis Methodology}
\label{sec.course.eval}

Rather than analyzing how a visualization corresponds to an instructor's grade, we focused on peer review feedback that students \textit{gave} to one another. In a prior paper, we found that students perceived the most benefit from reviewing others and providing feedback, rather than self-reflecting or receiving feedback from others~\cite{beasley2020leveraging}. Many students explicitly mentioned this aspect as most helpful in a post-course survey, and it is consistent with other researcher's findings~\cite{garousi2010applying}. By analyzing the reviews a student gave, we could better determine student engagement and the relationship between a student's comments and their mastery of concepts in the course.

\begin{figure}[!b]
    \centering
    \vspace{-10pt}
    \includegraphics[width=1\linewidth]{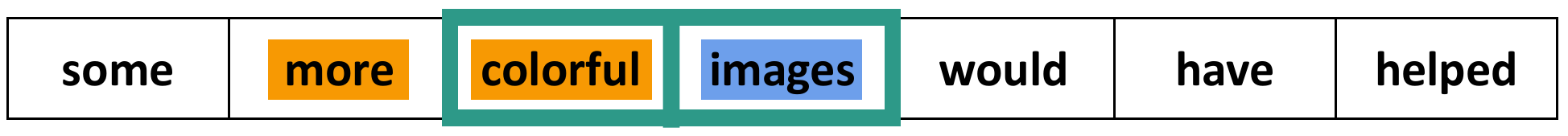}
    \vspace{-10pt}
    \caption{Aspect Extractor example where a noun appears in close proximity to an adjective.}
    \label{fig:ae}
\end{figure}

Textual feedback from both courses was analyzed with a dictionary-based natural language processing algorithm~\cite{beasley2021domain}. The algorithm matched positive and negative sentiment-bearing keywords to produce metrics that included overall sentiment of the review (positive or negative), counts of parts-of-speech (i.e., noun, adjective, adverb), and the average length of comments. 
The algorithm included an \textit{aspect extractor} that scanned text in a sliding window and produced a list of important aspects (nouns) in close proximity to sentiment words (adjectives). The aspects were frequently commented upon words associated with either positive or negative sentiment that indicating their importance to reviewers. We compared these words to the rubric to determine whether students stayed on topic or commented on unrelated topics. \autoref{fig:ae} shows, highlighted in teal, the adjective (``colorful'') and noun (``images'') matching within the sliding window. While ``colorful'' has a positive inferred sentiment, it is modified by the comparative adjective, ``more,'' which reverses the sentiment to negative. It is worth noting that the development of the algorithm targets peer review in engineering courses. It was not explicitly tuned for visualization courses or the particular rubric used. Thus, the algorithm did not ``look'' for visualization keywords, only for general nouns with associated sentiment-bearing words.

The field of writing composition provides a useful reference for our non-experimental evaluation methodology. Mulder et al., e.g., used similar qualitative and quantitative methods by performing content analysis on peer review comments and student questionnaires~\cite{Mulder2014casestudy}. Analyzing the words, both number and variety, used in written comments has also been used for measuring both learning outcomes~\cite{McGourty19989Stuentpeerreview} and student engagement~\cite{van2010effective}.

\section{PEER REVIEW AND STUDENT OUTCOMES}

Peer review textual feedback, though ripe with information, may largely be ignored in large courses. 
However, the unstructured responses can provide insights into course content, student engagement, and the peer review process as a whole to indicate students' visual literacy and the effectiveness of visualization education.

\subsection{Reinforce Course Content with Peer Review}
\label{sec.contents}

We began by analyzing the aspects (nouns) in student peer review comments to determine whether peer review reinforces course content in the visualization classroom. If students mentioned key concepts learned in the course in their rubric responses, we interpreted it to mean that they identified course content in context. 
%

\begin{figure}[!ht]
    \subfloat[Data Visualization aspects\label{fig:sentimentDV}]{%
        \includegraphics[trim=0 60pt 250pt 0, clip, width=1.0\linewidth]{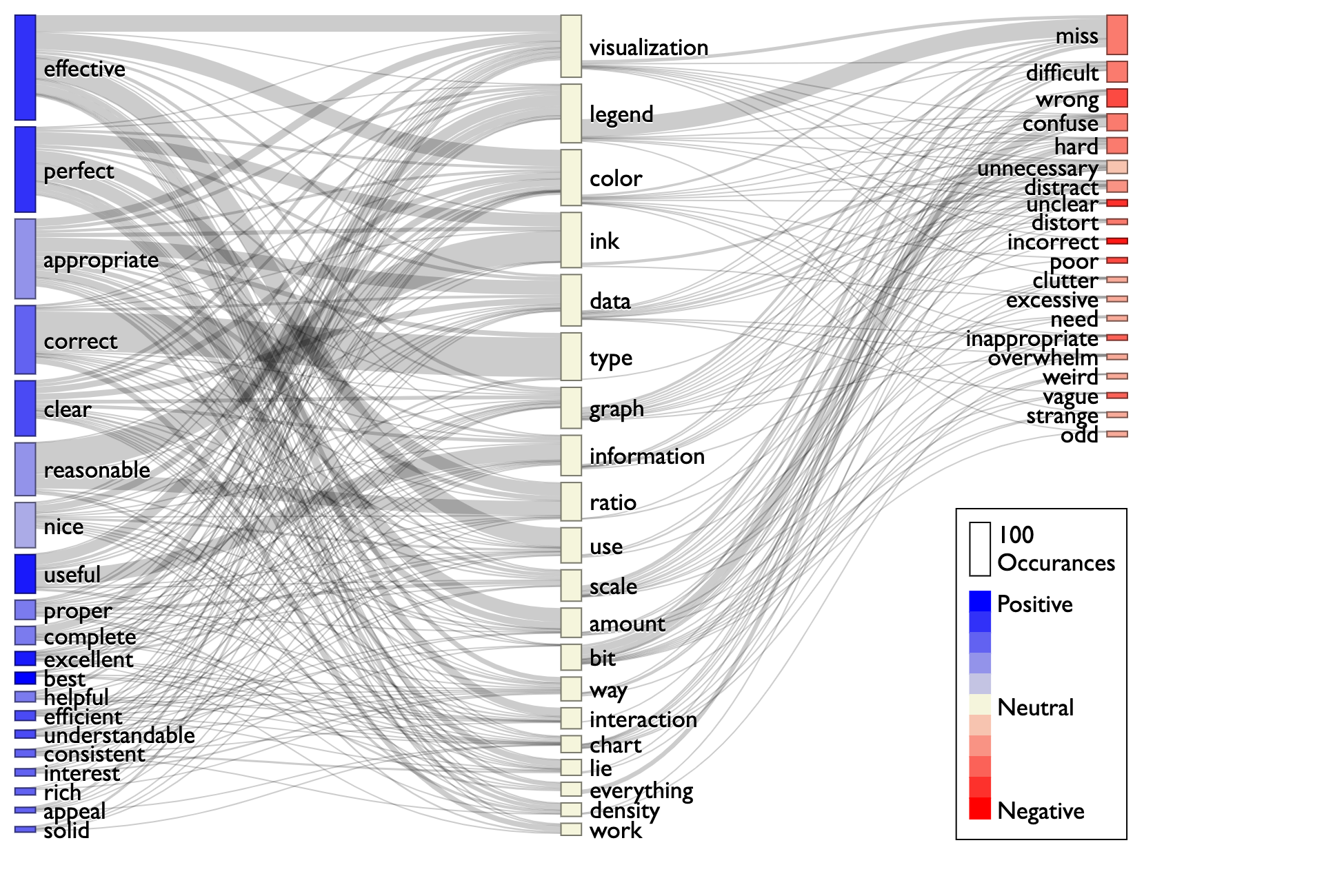}
    }
    \hfill
    \subfloat[Visual Literacy aspects\label{fig:sentimentVL}]{%
    \includegraphics[trim=0 60pt 280pt 0, clip, width=1.0\linewidth]{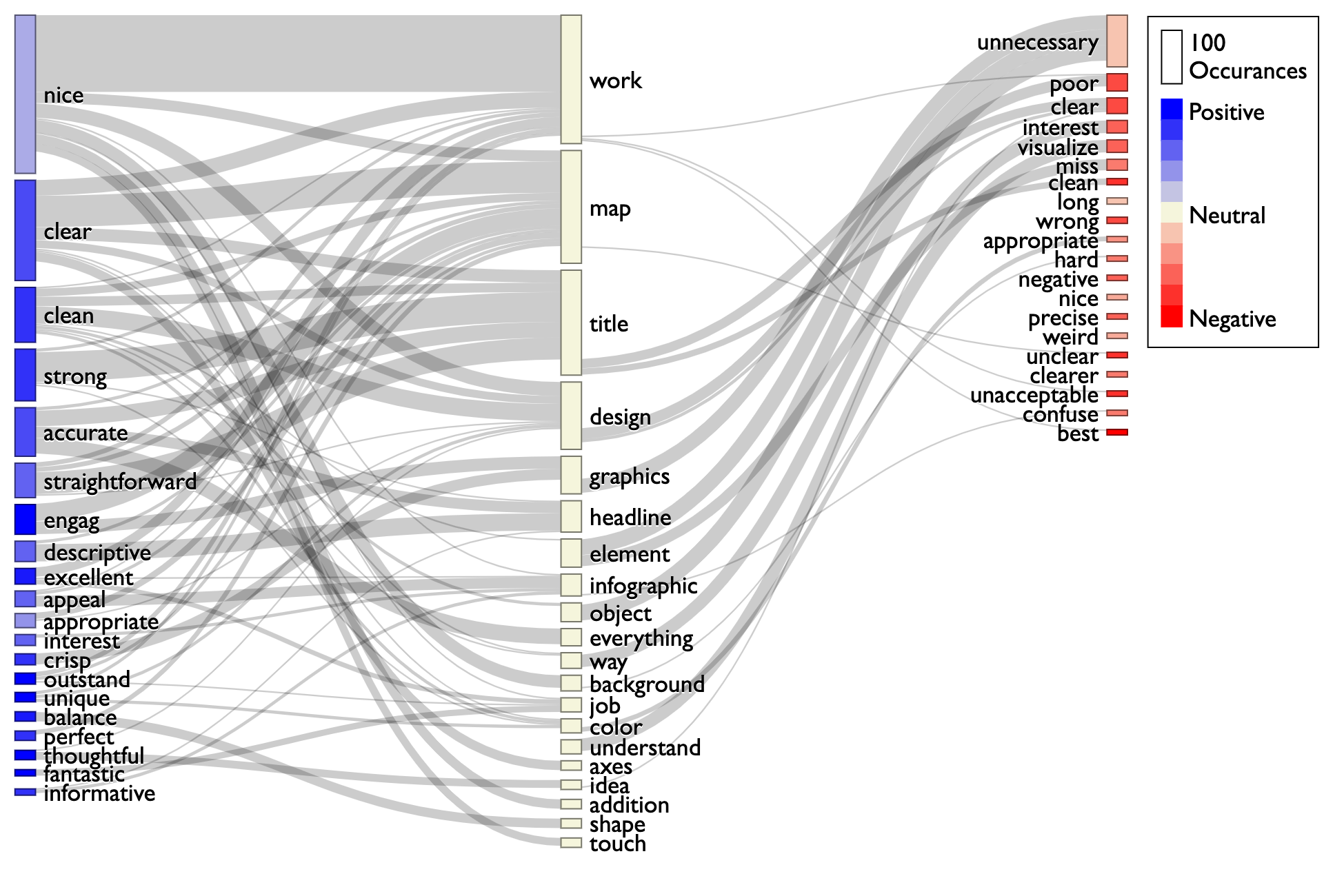}
    
    }
     \caption{Comparison of peer review sentiment. Aspects (center columns) are connected to positive (left) and negative (right) sentiment words that they appear by in text.}
    \label{fig:aspectComparison}
  \end{figure}

For the Data Visualization course, the top 20 aspects included: 
visualization, 
legend,
color,  
ink, 
data,
type (of data),  
graph, 
information, 
ratio (of data), 
use (of color, encoding), 
scale, 
amount (of ink, data), 
interaction, 
chart,  
lie, and 
density. Perhaps unsurprisingly, these words matched those of the rubric. \autoref{fig:sentimentDV} provides context. It displays the aspects in the center column with the number of connected positive and negative (or negated positive) sentiment words to the left and right columns, respectively. Saturation denotes the strength of sentiment, and the height of the bars indicates the number of occurrences.

\begin{table*}[!t]
    \centering 
    \caption{Analytic evaluation of peer comments in Data Visualization (DV) and Visual Literacy (VL)}
    \label{table:pcDV}
    
    \renewcommand{\arraystretch}{1.2}
    \resizebox{1.0\linewidth}{!}{%
        \begin{tabular}{lccccccccc}
         & \multirow{2}{*}{Reviews} & Avg & Avg Positive & Avg Negative & Avg Words  & Avg Words    & Avg        & Avg     & Avg \\
         &  & Keywords & Keywords & Keywords & Per Review & Per Sentence & Adjectives & Adverbs & Nouns \\
         \cline{1-10}
        DV Graduate & 1543 & 7.75 & 4.99 & 2.76 & 100.31 & 7.85 & 9.78 & 4.91 & 30.50 \\
        DV Undergraduate & 1573 & 9.26 & 5.78 & 3.48 & 135.61 & 10.45 & 12.42 & 8.37 & 36.86 \\
        \cline{1-10}
        VL Freshman&111&2.50&1.54&0.96&34.84&12.10&3.48&2.03&8.78\\
        VL Sophomore&109&2.59&1.72&0.87&38.26&14.58&3.74&2.28&9.28\\
        VL Junior&17&2.82&1.65&1.18&36.88&15.29&3.35&2.12&9.35\\
        VL Senior&879&1.89&1.28&0.61&22.96&11.14&2.50&1.16&6.50\\
        \cline{1-10}
        \end{tabular}
    }
    \renewcommand{\arraystretch}{1.0}
    
\end{table*}

In the Visual Literacy course, the top 20 aspects included:
work, 
map, 
title, 
design, 
graphics, 
headline, 
element, 
infographic, 
object, 
background (of visualization), 
understand (the visualization), 
color, 
shape, and  
axes.
Interestingly, these aspects (\autoref{fig:sentimentVL}) form a very different set than those in Data Visualization, which hints at a distinction between students and course content. Students did not often mention the language of the rubric, which included ``Detailed label,'' ``Lie factor,'' ``Data/color ink ratio,'' and ``Chart Junk.'' Students provided similar ratios of positive and negative feedback on their aspects to those in Data Visualization. Students also used positive and negative modifiers more consistently, showing more agreement in phrasing. When combined with other indicators of engagement, these students appear less likely to stay on topic and give thorough reviews.

While we cannot directly measure if students understand concepts better through the use of peer review, we can observe the approach providing repeated exposure to important concepts, both giving and receiving on-topic feedback. In addition, students are, at the very least, frequently repeating terminology in context, thereby increasing their visual literacy and critical analysis skills.

\subsection{Understand Student Engagement with Peer Review}


We next considered whether students engage in the peer review process. If not, the advantages of reinforcing course content are lost. As recommended by~\cite{van2010effective}, we analyzed the number and variety of words to quantitatively evaluate student engagement. We specifically identified nouns (aspects), adjectives (aspect modifiers), and adverbs (sentiment enhancers) in the peer review comments. The relevant summary statistics for both Data Visualization and Visual Literacy are shown in \autoref{table:pcDV}. 

We noticed that undergraduate students in Data Visualization wrote more than graduate students. In addition, they used more words per sentence, and they used a greater variety of tagged parts of speech, especially adverbs. Interestingly, undergraduates and graduates had a similar ratio of negative to total keywords, $38\%$ and $36\%$, respectively. This is an important measure of engagement because critically evaluating a visualization requires more investment than just a cursory review---it requires explaining \textit{why} something is wrong using learned concepts (e.g., analyzing for ``lie factor''). Considering factors, such as length, variety of parts of speech, and the ratio of negative keywords indicates that undergraduate students may be slightly more invested in the peer review process than their graduate peers.

In contrast to the Data Visualization course, peer review comments in Visual Literacy were from undergraduate only and were coded by class level. This enabled a more fine-grained analysis. It should be noted that juniors comprised a significantly reduced dataset (19) compared to freshmen (149), sophomores (141), and seniors (1262), so it is difficult to infer statistically significant results for those students.

\begin{figure}[!b]
    \centering
    {\includegraphics[width=1.0\linewidth]{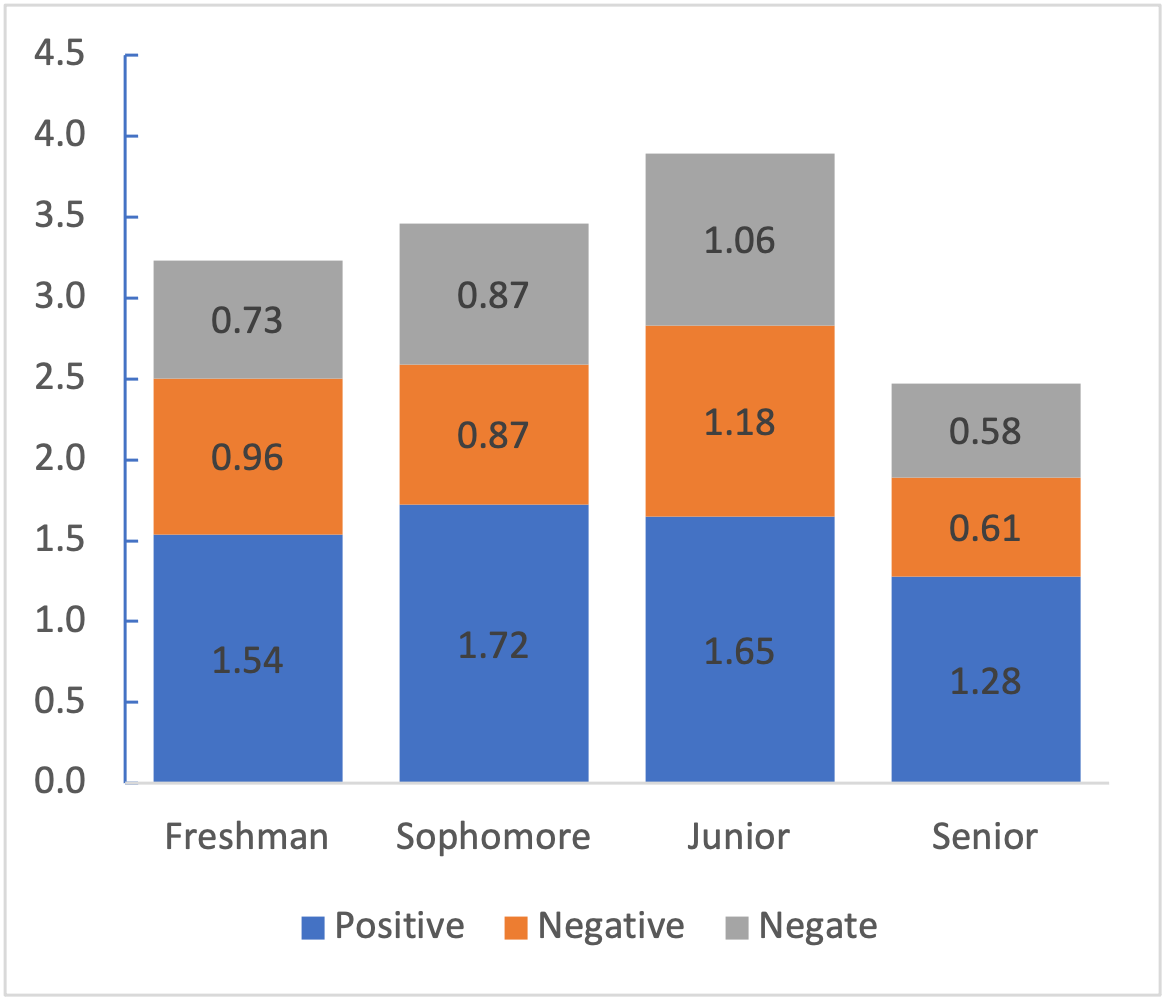}}
    
    \caption{Average Sentiment Keyword Usage Per Class Level in Visual Literacy}
    \label{fig:keyword_cl}
\end{figure}

For Visual Literacy students, the ratio of negative to total keywords was 33\%, which was lower than the Data Visualization students. When broken down by class level, we found juniors (42\%) were the most critical, followed by freshmen (37\%), sophomores (34\%), and finally seniors (32\%). Ultimately, senior students seem the least engaged through their comparative lack of use of sentiment-bearing keywords (\autoref{fig:keyword_cl}).

\begin{figure*}[!t]
	\centering
	\includegraphics[trim = 0 0 0 0, clip,width=1.0\linewidth]{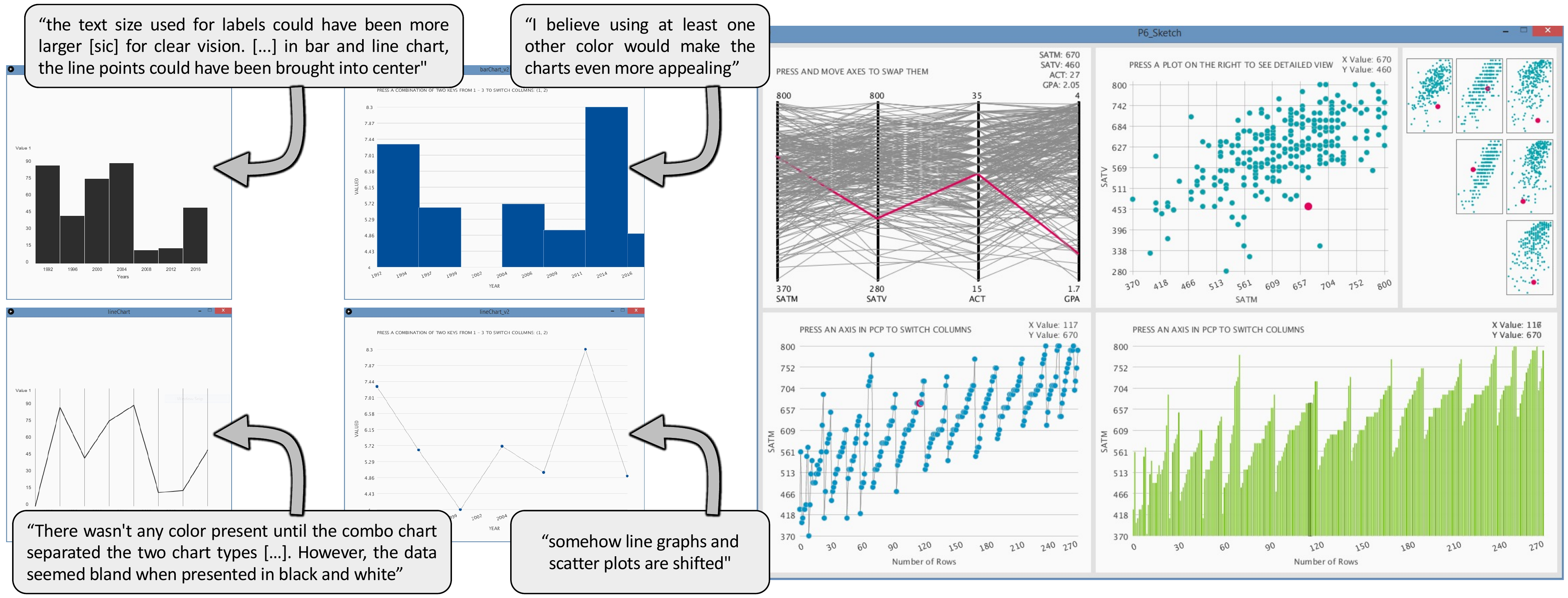}
	
	\begin{minipage}{0.20\linewidth}
    \end{minipage}
	\hspace{7pt}
	\begin{minipage}{0.2\linewidth}
    \end{minipage}
    \hfill
	\begin{minipage}{0.525\linewidth}
	\end{minipage}
	
	\vspace{-10pt}
	\caption{Example progression of a single student's three projects with associated feedback received and utilized to improve the design in Data Visualization.}
	\label{fig:studentFeedback}
	\vspace{-5pt}
\end{figure*}

The Visual Literacy student review comments were shorter than those of the Data Visualization students (average words per review), and they also exhibited less sentiment (average positive or negative keywords) and part-of-speech variety (average adjectives, adverbs, and nouns). Visual Literacy students did, however, write longer sentences, although they wrote fewer. The combination of this information can perhaps be attributed to the level of detail in the rubrics utilized. In Data Visualization, the rubric provided multiple textboxes for students to write a detailed analysis on each rubric item, rather than the single textbox of the Visual Literacy rubric.

The Data Visualization course provides an additional piece of evidence: a qualitative perspective on the engagement of students in the peer review process. It highlights the difference between submissions to demonstrate the effect of a student \textit{receiving and implementing} peer feedback on their iterative project.
In \autoref{fig:studentFeedback}, the initial bar and line charts received the comment: ``the text size used for labels could have been more larger [sic] for clear vision. [...] in bar and line chart, the line points could have been brought into center'' and ``There wasn't any color present until the combo chart separated the two chart types with a red color. However, the data seemed bland when presented in black and white.'' To respond to the comments, the student changed the bar chart color, centered the line chart points to the label (actually displaying the points themselves) and made the axis titles slightly larger, as shown in the center charts \autoref{fig:studentFeedback}.
The projects in the center column of \autoref{fig:studentFeedback} received comments, including: ``somehow line graphs and scatter plots are shifted'' and ``I believe using at least one other color would make the charts even more appealing.''  The student added another color to the dashboard (right chart in \autoref{fig:studentFeedback}) and shifted the appropriate graphs to not overlap in response. Therefore, in addition to the benefit of reviewing others' work, the student appeared to consider and implement a fair portion of the feedback they received.


\subsection{Mastery of Concepts and Peer Review}
\label{sec.performance}


Next, we considered whether student sentiment in comments \textit{provided} to peers was related to performance on the project, thereby establishing a link between engagement in peer review (sentiment or part-of-speech metrics) and visual literacy (grade on assignment). Although student grade information was not available in Data Visualization, students perceived benefit from peer review in general (and in particular, through \textit{providing} feedback to their peers), despite the extra time it took \cite{beasley2020leveraging}. Over three semesters, 82\% of post-course survey respondents reported learning at least somewhat more because of peer review (score of 3 out of 5 or more; mean~=~3.6) and 75\% recommended continuing peer review (score of 4 out of 5 or more; mean~=~4.1) \cite{beasley2020leveraging}.

\begin{figure}[!b]
    \centering
    {\includegraphics[width=1\linewidth]{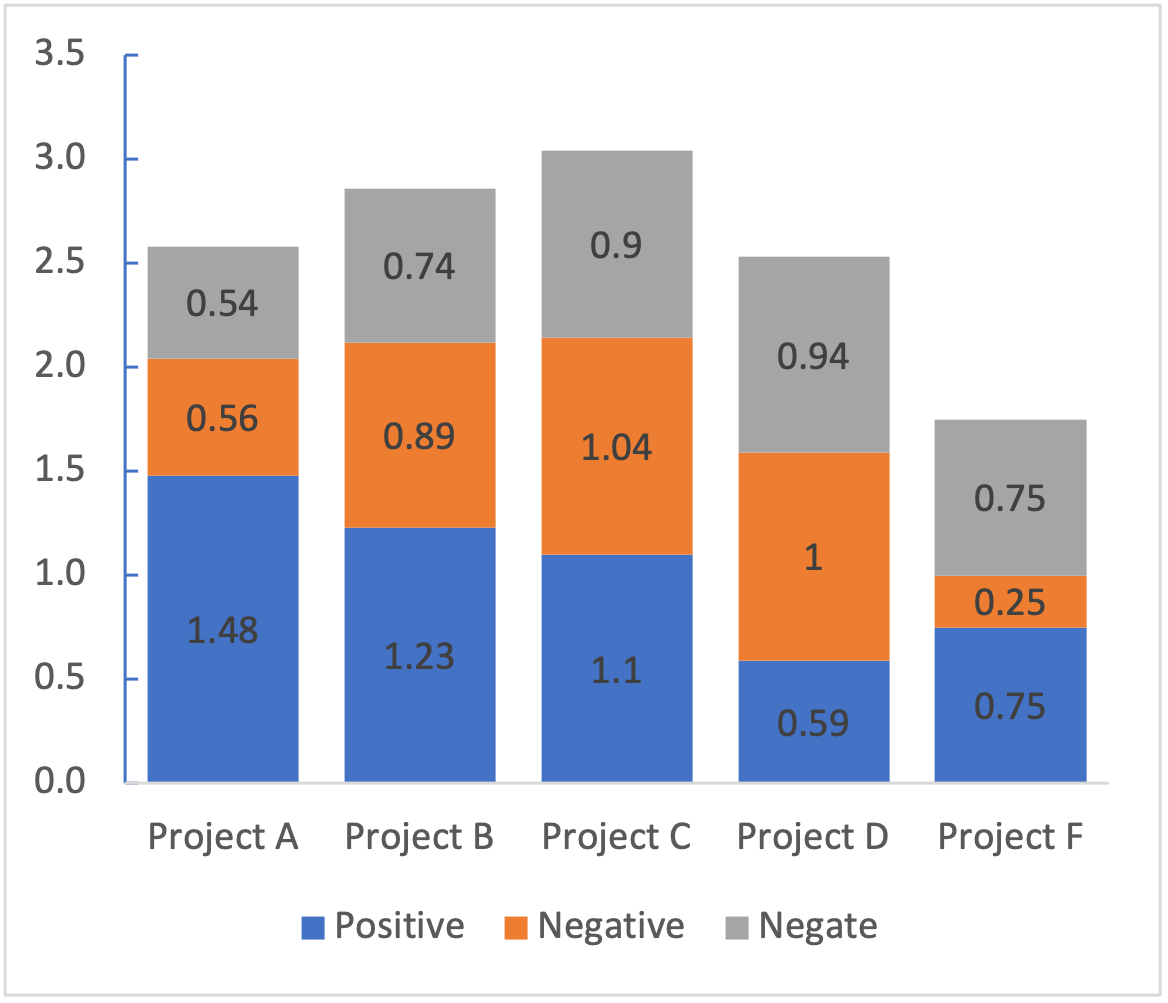}}
    
    \caption{Average Sentiment Keyword Usage Per Project Grade in Visual Literacy}
    \label{fig:keyword_pg}
\end{figure}

Grade information was available in Visual Literacy, and \autoref{fig:keyword_pg} shows the average number of keywords in students' reviews by their project grade. In total, 755 students earned an `A', 216 earned a `B', 78 earned a `C', 63 earned a `D', and 4 earned an `F'. Although we cannot determine that those who write short reviews will receive an `F' due to the small sample size, it is interesting to note that students earning a `C' wrote the most. This is perhaps because `A' students felt comfortable that they would do well in the course without devoting additional effort, while borderline students felt the need to improve their performance by devoting additional effort to the reviews.

\begin{figure}[!b]
    \centering
    {\includegraphics[width=1\linewidth]{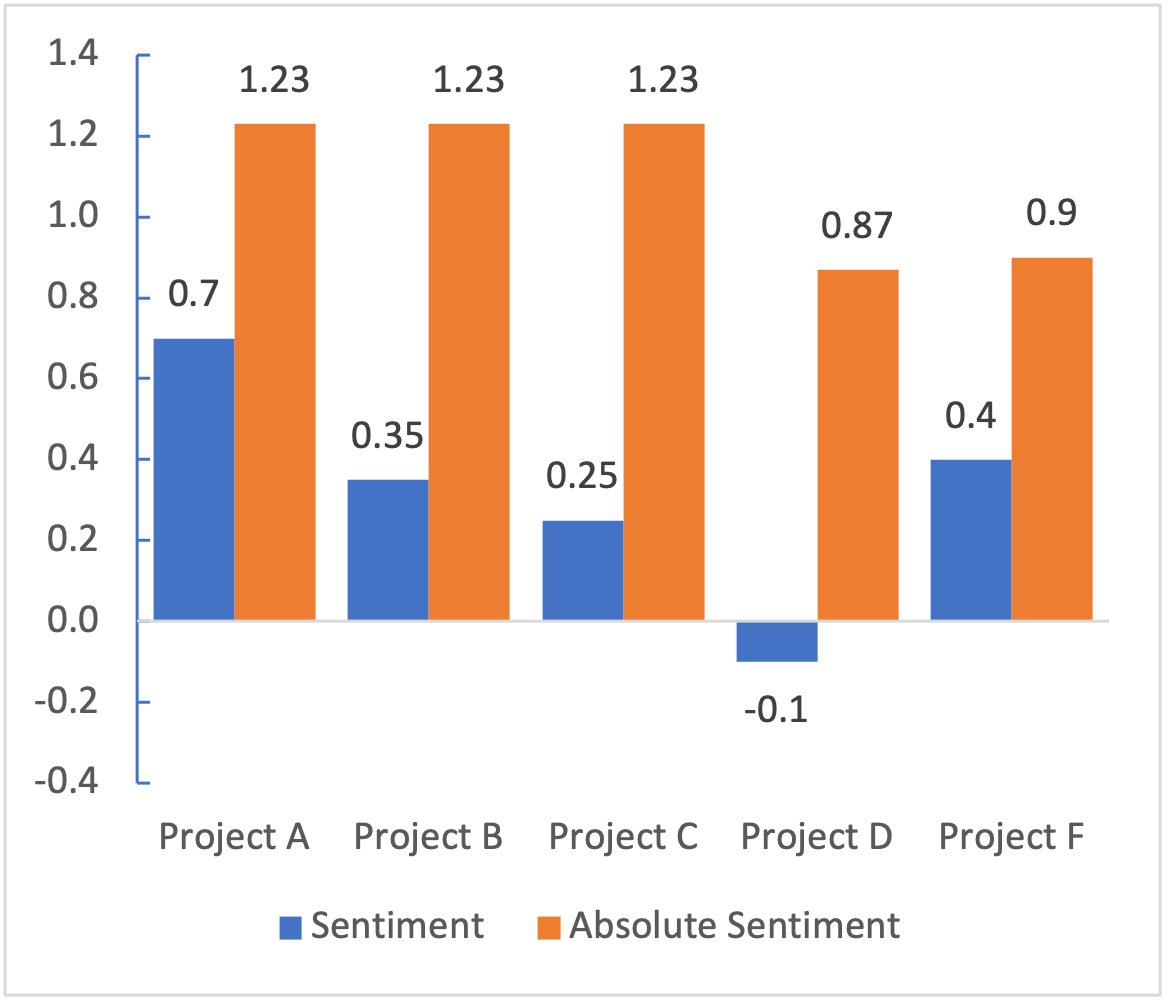}}
    
    \caption{Average Overall Sentiment and Absolute Value of Sentiment Per Project Grade in Visual Literacy}
    \label{fig:sentiment_pg}
\end{figure}

When comparing the sentiment used in reviews, there is one group of students that stands out in \autoref{fig:sentiment_pg}. Those earning a D on their project grade were the only group that wrote more negatively on average. Perhaps these critical students hoped to learn by carefully analyzing their peers' work for mistakes, or perhaps they (more pessimistically) desired to make their work look better in comparison.

\begin{figure*}[!t]
    \centering
    \includegraphics[width=1.0\linewidth]{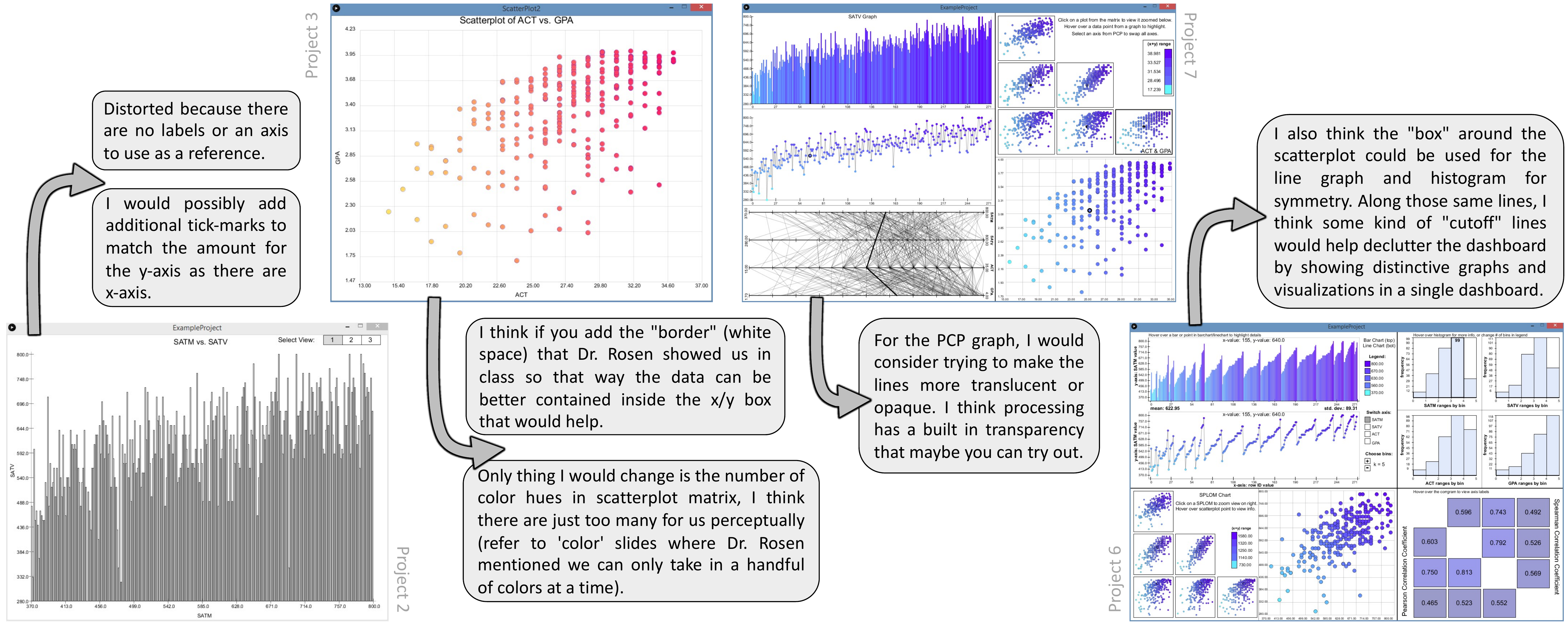}
    \caption{Illustrative example of a student's projects and the feedback they \textit{gave} to their peers, reflecting applied concepts in Data Visualization.}
    \label{fig:nofeedback}
    
\end{figure*}

One final qualitative observation from reviewing student work provides more evidence for the connection between student feedback and performance in the visualization classroom. In the Data Visualization course, we found that students often comment on others' work in ways that they have already implemented in their visualizations. \autoref{fig:nofeedback} shows both a student's projects and the feedback that the student \textit{gave} to peers. In Project 2, for example, the student mentions axis labels and ticks---something they had carefully implemented in their bar and line charts. The student pointed out the correct use of color, and they directly referenced materials learned in class in their Project~3 feedback. In their Project 6 feedback, the student referenced a specific technique for avoiding clutter in a parallel coordinates plot. Finally, in Project~7 feedback, they included a tip to delineate charts on the dashboard. In all situations, the student offers advice that directly corresponded to a technique they implemented. This example corroborates our finding that peer review reinforces course content by providing students an avenue to communicate recently learned concepts. This is an option they might not otherwise have. Thus, the mere opportunity of peer review may be a significant factor for student success.

\subsection{Measure Intervention Effectiveness}
The previous analysis has suggested, from student data, how sentiment analysis of peer review text can benefit visualization students and increase visual literacy. However, peer review data can also provide insights into course trends over time, establishing baselines and providing a metric for class success.

\begin{figure}[!b]
    \centering
    {\includegraphics[width=1.0\linewidth]{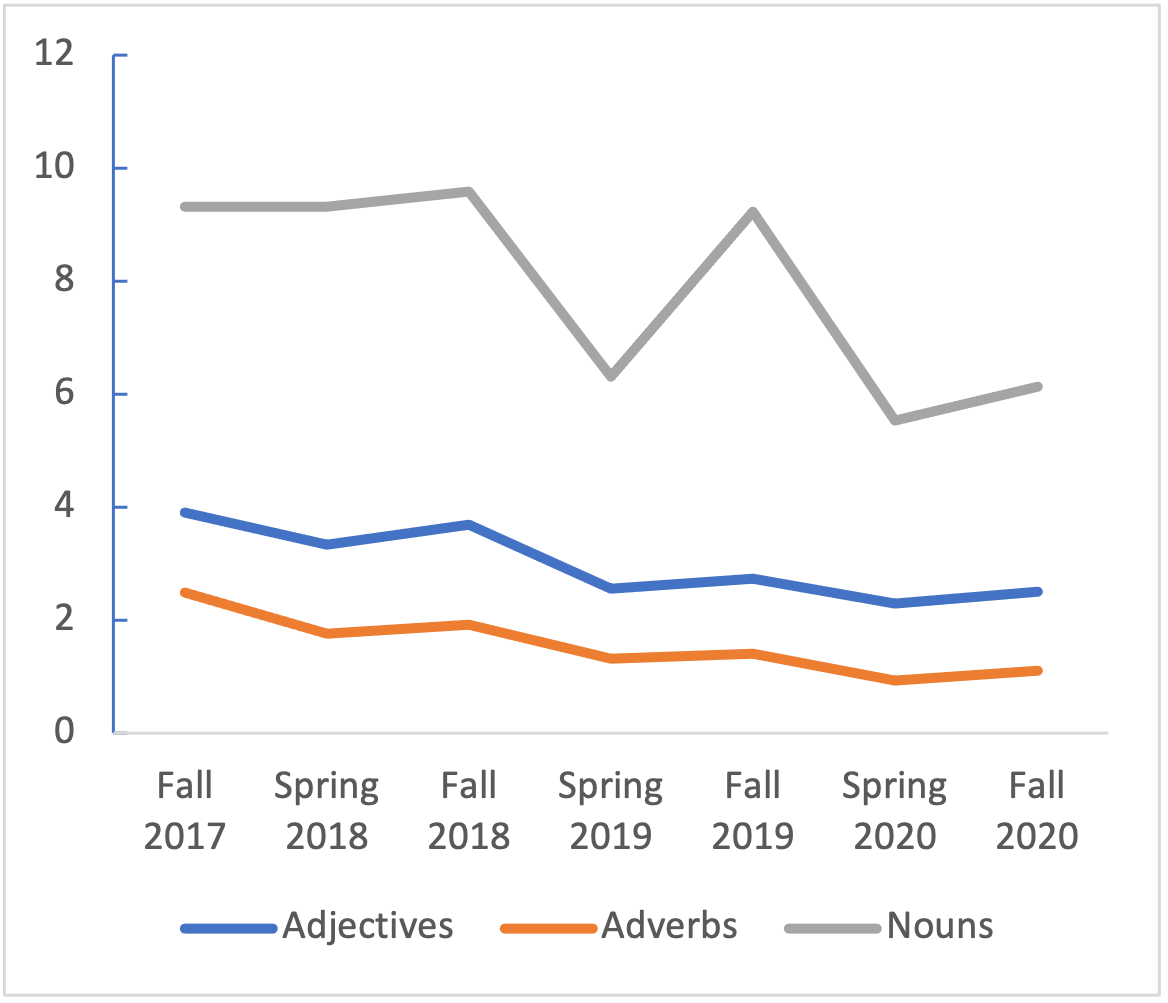}}
    
    \caption{Average Part-of-Speech Usage Per Review By Semester in Visual Literacy}
    \label{fig:pos_year}
\end{figure}

For example, \autoref{fig:pos_year} shows the variance in part-of-speech utilized by Visual Literacy students per semester. The trend lines highlight several interesting phenomena over the lifetime of the course. First, student part-of-speech variety has decreased on average from Fall 2017 to Fall 2020. As time progressed and students took multiple classes with the same instructors, they may have become more comfortable with the minimum level of expected effort. Since spring semesters are typically lower than fall semesters, it might also indicate burnout.

The major decrease from Fall 2018 to Spring 2019 corresponded to two factors in Visual Literacy: 1) a re-positioning of the course from senior-level to freshman-level and 2) a transition in modality from face-to-face to online in response to Covid-19. The combination of these factors may have contributed to the rapid decrease in the variety of part-of-speech usage.

When the instructors of the Visual Literacy course were questioned about factors related to the large increase from Spring 2019 to Fall 2019, they mentioned that in Fall 2019, USF provided expert support for the development of course materials and the improvement of assignments. In particular, the instructors were coached to employ verbs in their project and rubric descriptions. While this indicates the effectiveness of such expert intervention and underscores the importance of well-designed materials in a large, online course, ultimately, it appears the students quickly adjusted the next semester (Spring 2020) to a reduced level of engagement.

\subsection{Instructor Perspective: Observed Benefits}

We have observed several additional benefits from an instructor's perspective that we informally evaluate.

The teaching assistant for the 2018 and 2019 Data Visualization course pointed out that one advantage of peer review: formalizing existing collaboration. Many students discuss projects with one another and seek input without instructor intervention. However, peer review also initiated interaction for some students who would otherwise not interact with and offer constructive criticism to others. To them, it offered a lower risk environment for sharing their opinions than in class discussions. Requiring feedback in this way is important for the educational process by practicing critical evaluation skills, as well as to the visualization design process by providing a greater variety of perspectives.



Secondly, since many visualization courses cover specific programming languages required to generate visualizations, the instructor can include a code review phase that adds a new dimension to student engagement. This helps develop the students' ability to appraise their work from multiple perspectives (i.e., from peer review comments on the visualization's `front-end' to self-reflection on the code `back-end'). A student employing visual peer review in this manner is adding inherently asymmetrical analysis skills.

Finally, peer review feedback provides the instructor not only with a diverse set of opinions from which to review the visualization work but also provides insight into student learning from the quality of their review comments. This is yet another tool by which an instructor can identify successful students and is a tool for increasing transparency in the grading process.

\section{LIMITATIONS}

While the results analyzed above are highly specific to our courses, the approaches are by and large generalizable to other design-oriented visualization courses. 
If an instructor already uses a semester-long continuous design project, they may incorporate one or more peer review stages using our rubric or one of their own design. Beyond reinforcing concepts, a continuous improvement process is supported by the feedback given in peer review. If projects are discrete, the feedback cycle cannot be utilized, but many of the important peer review benefits will be retained, such as the reinforcement of course concepts or increased engagement.

\subsection{Instructor Effort}
On the surface, peer review appears to result in substantial time savings for instructors who no longer need to provide subjective feedback. For example, 60 students with eight assignments per semester and 10 minutes of feedback per assignment should result in over 80 hours saved. However, much of that time is diverted to other activities. Additional administrative tasks arise---peer review requires additional effort to set up and administer, peer reviews need to be monitored for quality, and grades need to be assigned. At the same time, highly engaged students might request more detailed instructor feedback beyond peer reviews. Overall, peer review is a compliment, not a replacement, to existing educational approaches that requires the instructor, as the expert, to remain involved.


\subsection{Risks}

When considering the use of peer review, the risk for collusion, malice, and cheating must be considered, even with double-blind reviews (e.g., the review of a friend may be overly optimistic). Instructors should be careful to randomize peer reviewers. Keeping submissions anonymous is helpful, but it is challenging to maintain anonymity, particularly in a smaller educational setting where students will talk and discover they are evaluating each other’s work. In conjunction, grades should only be loosely based upon the results of the peer review.


\section{CONCLUSION}

In this work, we contrast peer review in two visualization courses with distinct cultures and domains to identify the variety of information that can be obtained from the analysis of peer review text. We discovered differences between the two domains regarding the taxonomy of student comments to their peers. However, we also found similarities regarding student engagement and reinforcement of the core concepts each instructor presented in class. Sentiment analysis and aspect extraction of peer review comments augment the instructor's ability to draw insights on student engagement and success in the classroom, the focus of course content, course baselines, and the effectiveness of instructor intervention. 

Peer review in the visualization classroom offers two significant benefits to students. First, it provides a mechanism for students to actively engage through critical evaluation of visualizations, increasing visual literacy and awareness. Second, it enables providing students with diverse and timely feedback on their projects and other coursework. 
Ultimately, we encourage the visualization community to adopt peer review, with its rubrics and associated analysis, to improve instructor understanding of student engagement and promote visual literacy through critical analysis. 

In the future, we are interested in the customization of the rubric by reducing constraints. Assuming students' critical thinking skills are underdeveloped, particularly in the domain of visualization education, the rubric serves as an important tool to improve those skills. At the beginning of a course, the rubric can include all relevant scoring categories, and as the course progresses, related categories can be combined or removed. 
This way, students will progress from highly structured evaluation to entirely free-form evaluation.

Secondly, we noted a lack of high-quality reviews. Approximately 30\% of reviews in each class did not contain enough information to receive a fine-grained sentiment score from the natural language processing algorithm. While less critical for part-of-speech analysis of engagement, the addition of a fine-grained sentiment score on peer review text can provide a complementary component to the numerical score from the review form.
Measuring and improving the `helpfulness' of peer reviews with real-time suggestions for students with weak responses is an active area of research that we would like to incorporate into our future courses. This can be done by gamifying peer review to include a top reviewers scoreboard.

Finally, it is important to remember that there is both art and science to visualization, with no single optimal design. Using the wisdom of the crowd to build machine learning models that leverage peer feedback to semi-automatically assign grades is an exciting direction for future study.  


\paragraph{Disclosure} While the first author (Beasley) was a graduate student in the 2017 Data Visualization course, his participation in the research began after the course was completed.

\section{ACKNOWLEDGMENT}

We thank our anonymous reviewers for their helpful feedback on our paper. We also thank Coby O'Brien and Kevin Hawley for supporting our effort to promote visual peer review in their classrooms and Ghulam Quadri for his perspective as a teaching assistant in the Data Visualization course. This project was supported in part by the National Science Foundation (IIS-1845204).

\bibliographystyle{abbrv-doi-narrow}
\bibliography{main}

\begin{IEEEbiography}{Zachariah J. Beasley}{\,}is an Instructor 1 at the University of South Florida in the Department of Computer Science, where he received his Ph.D. with a focus on sentiment analysis in peer review. Dr. Beasley has received the ASEE State of Engineering Education in 25 Years Award, the USF Spirit of Innovation Award, and is a USF STEER STEM Scholar. Contact him at zjb@usf.edu.
\end{IEEEbiography}

\begin{IEEEbiography}{Alon Friedman}{\,}is an Associate Professor at the University of South Florida in the School of Information. He received his Ph.D. from the Palmer school of library \& information science at Long Island University. His recent research interests include using different theoretical perspectives to organize data through a visualization lens. His current projects include Mark Lombardi's design, Peircean semiotics theory through visualization lens, and Visual Peer review. Contact him at alonfriedman@usf.edu.
\end{IEEEbiography}

\begin{IEEEbiography}{Paul Rosen}{\,}is an Assistant Professor of Computer Science at the University of South Florida. His Ph.D.\ is from Purdue University. His recent research interests include geometry- and topology-based problem-solving in visualization. He and his collaborators have received several best paper awards and honorable mentions, and he received a National Science Foundation CAREER Award in 2019. Contact him at prosen@usf.edu.
\end{IEEEbiography}

\end{document}